\documentstyle[multicol,aps,epsfig]{revtex}

\tighten       

\begin{document}
\draft 

\title{Wavefront depinning transition in
discrete one-dimensional reaction-diffusion
systems}
\author{A. Carpio$^1$ and L. L. Bonilla$^2$ }
\address{$^1$Departamento de Matem\'{a}tica
Aplicada, Universidad Complutense de Madrid,\\
28040 Madrid, Spain\\
$^2$Departamento de Matem\'{a}ticas, Escuela
Polit\'ecnica Superior,  Universidad Carlos III
de Madrid,\\ 
Avenida de la Universidad 30, 28911 Legan{\'e}s,
Spain }
\date{ \today  }
\maketitle

\begin{abstract}
Pinning and depinning of wavefronts are ubiquitous
features of spatially discrete systems describing
a host of phenomena in physics, biology, etc. A
large class of discrete systems is described by
overdamped chains of nonlinear oscillators with
nearest-neighbor coupling and controlled by
constant external forces. A theory of the
depinning transition for these systems, including
scaling laws and asymptotics of wavefronts, is
presented and confirmed by numerical calculations.
\end{abstract}

\begin{multicols}{2}
\narrowtext
Spatially discrete systems describe physical
reality in many different fields: propagation of
nerve impulses along mielinated fibers
\cite{kee87,kee98}, pulse propagation through
cardiac cells \cite{kee98}, calcium release waves
in living cells \cite{bug97}, sliding of charge
density waves \cite{cdw}, superconductor Josephson
array junctions \cite{jj}, motion of dislocations
in crystals \cite{nab67}, atoms adsorbed on a
periodic substrate \cite{cha95}, arrays of
coupled diode resonators \cite{diode}, and weakly
coupled semiconductor superlattices
\cite{bon94,car00}. A distinctive feature of
discrete systems (not shared by continuous ones)
is the phenomenon of wavefront pinning: for
values of a control parameter in a certain
interval, wavefronts joining two different
constant states fail to propagate \cite{kee98}.
When the control parameter surpasses a threshold,
the wavefront depins and starts moving
\cite{kee87,cdw,nab67,car00}. The existence of
such thresholds is thought to be an intrinsecally
discrete fact, which is lost in continuum
aproximations. The characterization of propagation
failure and front depinning in discrete systems
is thus an important problem, which is still
poorly understood despite the numerous inroads
made in the literature
\cite{kee87,bug97,cdw,nab67,kla00,mit98,kin01}.

\begin{figure}
\begin{center}
\includegraphics[width=8cm]{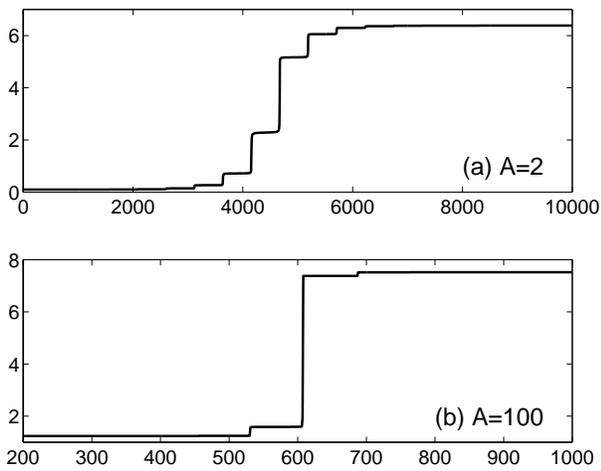}
\caption{Traveling wavefront profiles near $F=
F_c$ for the FK potential and: (a) $A=2$, (b)
$A=100$.}
\label{fig1}
\end{center}
\end{figure}

In this Letter, we study front depinning for
infinite one-dimensional nonlinear spatially
discrete reaction-diffusion (RD) systems. The
nature of the depinning transition depends on the
nonlinearity of the model, and is best understood
as propagation failure of the traveling front.
Usually, but not always, the wavefront profiles
become less smooth as a parameter $F$ (external
field) decreases. They become {\em discontinuous}
at a critical value $F_c$. Below $F_c$, the front
is pinned at discrete positions corresponding to a
stable steady state. Fig. \ref{fig1} shows
wavefront profiles near the critical field.
Individual points undergo abrupt jumps at
particular times, which gives the misleading
impression that the motion of the discrete fronts
proceeds by successive jumps. Wavefront velocity
scales with the field as $|F-F_c|^{{1\over 2}}$.
For exceptional nonlinearities, the wavefront
does not lose continuity as the field decreases.
In this case, there is a continuous transition
between wavefronts moving to the left for
$F>0$ and moving to the right for $F<0$: as for
continuous systems, front pinning occurs only at
a single field $F=0$. Wavefront velocity scales
then linearly with the field. We discuss the
characterization of the critical field (including
analytical formulas in the strongly discrete
limit), describe depinning anomalies (discrete
systems having zero critical field), and give a
precise characterization of stationary and moving
fronts near depinning (including front velocity)
by singular perturbation methods. Our
approximations show excellent agreement with
numerical calculations.

We consider chains of diffusively coupled
overdamped oscillators in a  potential $V$,
subject to a constant external force $F$:
\begin{equation}
{du_{n}\over dt} = u_{n+1}-2u_n + u_{n-1} + F -
A\,  g(u_n).   \label{Fd}
\end{equation}
Here $g(u)=V'(u)$ presents a `cubic' nonlinearity,
such that $A\, g(u)-F$ has three zeros, $U_1<U_2
<U_3$ in a certain force interval ($g'(U_i(F/A))
>0$ for $i=1,3$, $g'(U_2(F/A))<0$). Provided
$g(u)$ is odd with respect to $U_2(0)$, there is
a symmetric interval $|F|\leq F_c$ where the
wavefronts joining the stable zeros $U_1(F/A)$ and
$U_3(F/A)$ are pinned. For $|F|>F_c$, there are
{\em smooth traveling wavefronts}, $u_n(t)=
u(n-ct)$, with $u(-\infty)= U_1$ and $u(\infty)
=U_3$ \cite{zin92,car99}. The velocity $c(A,F)$
depends on $A$ and $F$ and it satisfies $cF<0$
and $c\to 0$ as $|F|\to F_c$ \cite{car99}.
Examples are the overdamped Frenkel-Kontorova
(FK) model ($g=\sin u$) \cite{FK} and the quartic
double well potential ($V=(u^2-1)^2 /4$). Less
symmetric nonlinearities yield a non-symmetric
pinning interval and our analysis applies to them
with trivial modifications.

{\em Critical field}. Stationary and traveling
wavefronts cannot coexist for the same value of
$F$ \cite{car99}. This follows from a comparison
principle for (\ref{Fd}) \cite{comparison}.
Pinning can be proved using stationary sub and
supersolutions, which can be constructed provided
the stationary solution is linearly stable. The
largest eigenvalue of the linearization of
(\ref{Fd}) about a stationary profile $u_n(A,F)$,
$u_n(t)= u_n(A,F) + v_n e^{\lambda t}$, is given
by
\begin{eqnarray}
-\lambda(A,F)=\mbox{Min}\, {\sum
(v_{n+1}-v_n)^2 +A g'(u_n(A,F)) v_n^2 \over \sum
v_n^2}\,, \label{var}
\end{eqnarray}
over a set of functions $v_n$ which decay
exponentially as $n\to\pm\infty$. The critical
field is uniquely characterized by $\lambda(A,F_c)
=0$ and $\lambda(A,F)<0$ for $|F| <F_c$. Thus two
facts distinguish the depinning transition: (i)
one eigenvalue becomes zero, and (ii) stationary
and moving wavefronts cannot coexist for the same
values of the field.

Equation (\ref{var}) shows that the critical field
is positive for large $A$ and typical
nonlinearities. In fact, consider the FK
potential. For $F=0$ there are two stationary
solutions which are symmetric with respect to
$U_2$, one taking on the value $U_2$ (unstable
dislocation), and the other one having $u_n \neq
U_2$ (stable dislocation) \cite{hob65}. For large
$A$, the stable dislocation has $g'(u_n)>0$ for
all $n$, and (\ref{var}) gives $\lambda(A,0)<0$.
Since $\lambda(A,F_c)=0$, this implies that the
critical field is nonzero. (A different proof can
be obtained using the comparison principle
\cite{kee87,car00}). As $A>0$ decreases, several
$u_n$ may enter the region of negative slope
$g'(u)$: the number of points with $g'(u_n)<0$
increases as $A$ decreases. It is then possible
to have $\lambda(A,0)=0$, i.e.\ $F_c=0$, for a
discrete system! Examples of this {\em pinning
anomaly} will be given below.

If $F>0$, the stable dislocation is no longer
symmetric with respect to $U_2$. If $F$ is not
too large, all $u_n(A,F)$ avoid the region of
negative slope $g'(u)<0$. For large $A$ and $F$
and the generic potentials above mentioned, we
have observed numerically that $g'<0$ for a single
point, labelled $u_0(A,F)$. This property
persists until $F_c$ is reached. How does $F_c$
depend on $A$? For $g=\sin u$, it is well known
that $F_c$ vanishes exponentially fast as $A$
goes to zero (the continuum limit). This was
conjectured by Indenbom \cite{ind58} on the basis
of a continuum approximation, and numerically
checked by Hobart \cite{hob65} in the context of
the Peierls stress and energy for dislocations.
More recently, Kladko et al
\cite{kla00} derived the formula $F_c =
C\,\exp(-\pi^2/\sqrt{A - A^2 /12})$ by means of a
variational argument. This argument can be used
for other potentials and it suggests that
$F_c\sim C\, e^{-\eta/\sqrt{A}}$ as
$A\to 0+$ (with positive $C$ and $\eta$
independent of $A$) holds for a large class of
nonlinearities \cite{kla00}. King and Chapman
have obtained an analogous result \cite{kin01}
using exponential asymptotics for the FK
potential, $F_c \sim C\, e^{-\pi^2/
\sqrt{A}}$ and the wavefront velocity after
depinning, $c\sim D\,\sqrt{(F^2- F_c^2)/A}$. This
later result agrees with the scaling law $c
\sim |F-F_c|^{{1\over 2}}$, found in a large class
of discrete RD equations \cite{kla00,mit98}.

{\em Anomalies of pinning}. Despite widespread
belief, it is not true that $F_c>0$ for all
discrete systems. Using the characterization
$\lambda(A,F_c)=0$, it is possible to see that
having a zero critical field is equivalent to
having a one-parameter family of continuous
increasing stationary profiles $u_n = u(n+\alpha)$
satisfying $u(x+1)+u(x-1)-2u(x) = A g(u(x))$,
with $u(-\infty)=U_1$, $u(\infty)=U_3$. In this
case, a standard perturbation argument yields a
wavefront speed having the same scaling as the
continuum approximation to the discrete system,
$c\sim C\, F$ as $F\to 0$. An example of
nonlinearity presenting this anomalous pinning
\cite{mcleod} can be obtained from $u(x)=\tanh
x$: it obeys the above equation with $A=1$, $U_1=
-1$, $U_3=1$ and $g(u)= - 2 \tanh^2(1)\, u
(1-u^2)/[1-\tanh^2(1)\, u^2]$. Furthermore the
wavefront velocity after depinning obeys the
relation, $c\sim - 3 F/2$ as $F\to 0$. Thus
nonlinearities presenting anomalous depinning
belong to a different universality class: the
wavefront velocity has a critical exponent 1 (and
$F_c=0$) instead of 1/2, which is the usual case
for discrete RD systems (having $F_c>0$).

\begin{figure}
\begin{center}
\includegraphics[width=8cm]{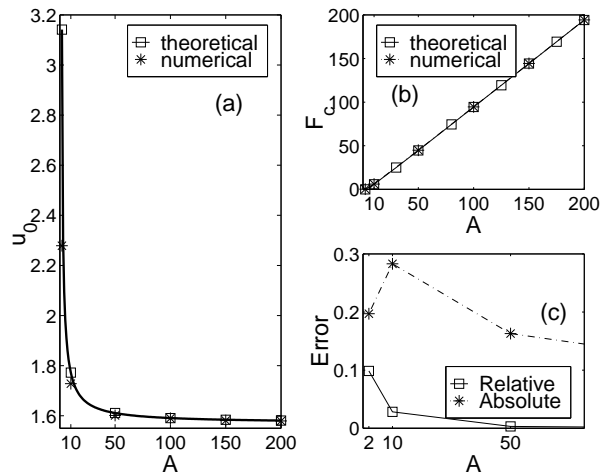}
\caption{(a) $u_0(A,F_c)$; (b) Critical field as
a function of $A$; (c) Absolute and relative
errors in $F_c(A)$.}
\label{fig2}
\end{center}
\end{figure}

{\em Asymptotic theory of wavefront depinning}.
We shall study the depinning transition in the
strongly discrete limit $A\gg 1$, in which the
structure of the wavefront is particularly
simple. Firstly, consider the symmetric
stationary profile with $u_n\neq U_2$ for $F=0$.
The front profile consists of two tails with
points very close to
$U_1$ and $U_3$, plus two symmetric points $u_0$,
$u_1$ in the gap region between $U_1$ and $U_3$.
As $F>0$ increases, this profile changes slightly:
the two tails are still very close to $U_1(F/A)$
and $U_3(F/A)$. As for the two middle points,
$u_1$ gets closer and closer to $U_3$ whereas
$u_0$ moves away from $U_1$. This structure is
preserved by the traveling fronts above the
critical field: there is only one active point
most of the time, which we can adopt as our
$u_0$. Then we can approximate $u_{-1}\sim U_1$,
$u_1\sim U_3$ in (\ref{Fd}), thereby obtaining
\begin{equation}
{du_{0}\over dt}\approx U_1\left({F\over A}
\right) + U_3\left({F\over A} \right)-2 u_0 -
A\, g(u_0) + F .\label{u0}
\end{equation}
This equation has three stationary solutions for
$F<F_c$, two stable and one unstable, and only one
stable stationary solution for $F>F_c$. The
critical field $F_c$ is such that the expansion
of the right hand side of (\ref{u0}) about the
two coalescing stationary solutions has zero
linear term, $2 +A g'(u_0)=0$, and
\begin{equation}
2 u_0 + A \, g(u_0) \sim  U_1\left({F_c\over A}
\right) + U_3\left({F_c\over A} \right) + F_c .
\label{Fc1}
\end{equation}
These equations for $F_c$ and $u_0(A,F_c)$ have
been solved for the FK potential, for which $u_0 =
\cos^{-1}(-2/A)$ and $U_1+U_3 = 2\sin^{-1}
(F_c/A) + 2\pi$. The results are depicted in Fig.
\ref{fig2}, and show excellent agreement with
the numerical solution of (\ref{Fd})
for $A>10$. Our approximation performs less well
for smaller $A$, and it breaks down at $A=2$ with
the prediction $F_c=0$. Notice that $F_c(A)\sim
A$ as $A$ increases. In practice, only steady
solutions are observed for very large $A$.

\begin{figure}
\begin{center}
\includegraphics[width=8cm]{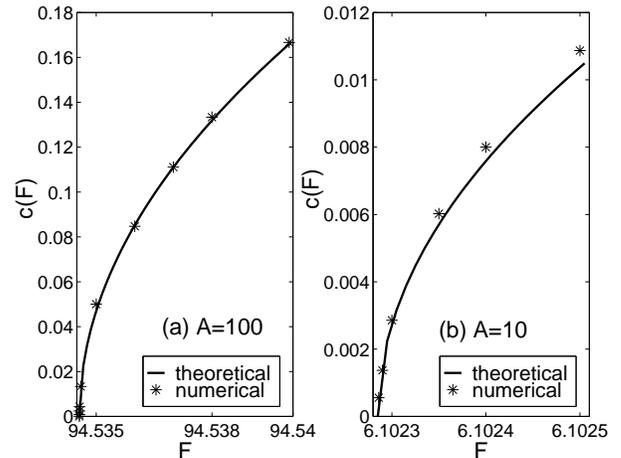}
\caption{Wavefront velocity versus $F$
near $F= F_c$ for the FK potential and: (a)
$A=100$, (b) $A=10$.}
\label{fig3}
\end{center}
\end{figure}

Let us now construct the profile of the traveling
wavefronts after depinning, for $F$ sligthly above
$F_c$. Then $u_0(t)= u_0(A,F_c) + v_0(t)$ obeys
the following equation:
\begin{equation}
{dv_{0}\over dt}\approx (F-F_c) + A\, |g''(u_0)|\,
{v_{0}^{2}\over 2} ,\label{v0}
\end{equation}
where we have used $2+Ag'(u_0)=0$, (\ref{Fc1})
and ignored terms of order $(F-F_c)/A$ and higher.
This equation has the (outer) solution
\begin{equation}
v_0(t)\sim \sqrt{{2\, (F-F_{c})\over A\, |g''(
u_{0})|}}\, \tan\left(\sqrt{{A\, |g''(u_0)|\,
(F-F_{c})\over 2}}\, (t-t_0)\right)\,,
\label{outer}
\end{equation}
which is very small most of the time, but it
blows up when the argument of the tangent function
approaches $\pm \pi/2$. Thus the outer
approximation holds over a time interval
$(t-t_0)\sim \pi\sqrt{2}/\sqrt{A |g''(u_0)|\,
(F-F_{c})}$, which equals $\pi\sqrt{2} (A^2-4)^{
-{1\over 4}} (F-F_c)^{-{1\over 2}}$ for the FK
potential. The reciprocal of this time interval
yields an approximation for the wavefront
velocity,
\begin{equation}
|c(A,F)|\sim \sqrt{{A\, |g''(u_{0})|\,
(F-F_{c})\over 2\pi^{2}}}\,, \label{c}
\end{equation}
or $|c|\sim (A^2-4)^{{1\over 4}} (F-F_c)^{{1\over
2}} /(\pi\sqrt{2})$ for a FK potential. In Fig.\
\ref{fig3} we compare this approximation with the
numerically computed velocity for $A=100$ and
$A=10$.

\begin{figure}
\begin{center}
\includegraphics[width=8cm]{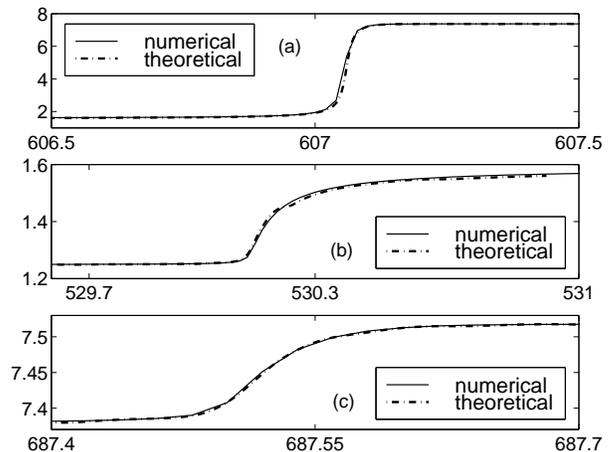}
\caption{Wavefront profiles near $F_c$ for the
FK potential and $A=100$. We show the three
largest jumps in Fig. \ref{fig1}(b). }
\label{fig4}
\end{center}
\end{figure}

When the solution begins to blow up, the outer
solution (\ref{outer}) is no longer a good
approximation, for $u_0(t)$ departs from the
stationary value $u_0(A,F_c)$. We must go back
to (\ref{u0}) and obtain an inner approximation
to this equation. As $F$ is close to $F_c$ and
$u_0(t)-u_0(A,F_c)$ is of order 1, we solve
numerically (\ref{u0}) at $F=F_c$ with the
matching condition that $u_0(t)-u_0(A,F_c)\sim
2/[\pi \sqrt{{1\over 2}\, A|g''(u_0)|/(F-F_c)} -A
|g''(u_0)|\, (t-t_0)]$, as $(t-t_0)\to -\infty$.
This inner solution describes the jump of $u_0$ to
values close to $U_3$. During this jump, the
motion of $u_0$ forces the other points to move.
Thus, $u_{-1}(t)$ can be calculated by using the
inner solution in (\ref{Fd}) for $u_0$, with
$F=F_c$ and $u_{-2}\approx U_1$. A composite
expansion \cite{bon87} constructed with these
inner and outer solutions is compared to the
numerical solution of (\ref{Fd}) in Fig.
\ref{fig4}.

Notice that (\ref{v0}) is the normal form
associated with a saddle-node bifurcation in a
one dimensional phase space. The wavefront
depinning transition is a {\em global} bifurcation
with generic features: each individual point
$u_n(t)$ spends a long time, which scales as
$|F-F_c|^{-{1\over 2}}$, near discrete values
$u_n(A,F_c)$, and then jumps to the next discrete
value on a time scale of order 1. The traveling
wave ceases to exist for
$F\leq F_c$. For these field values, discrete
stationary profiles $u_n(A,F)$ are found. The
above calculations give a normal form of the type
$d^2 v_0/dt^2 = \alpha\, (F-F_c) + \beta\, v_0^2$
instead of (\ref{v0}) for conservative discrete
systems (two time derivatives instead of one in
(\ref{Fd})). The solution of this equation blows
up in finite time as $(F-F_c)^{ -{1\over 4}}$,
which gives a critical exponent of 1/4 for the
wavefront velocity near the critical field.

The approximations to $F_c(A)$ and the wavefront
speed provided by the previous asymptotic theory
break down for small $A$. In particular, for the
FK potential and $A<2$, no double zeroes of $2x
+A\sin(x) - (F +U_{1} +U_3)$ are found for
$F=F_c$. What happens is that we need more than
one point to approximate wavefront motion.
Depinning is then described by a reduced system
of more than one degree of freedom corresponding
to active points. There is a saddle-node
bifurcation in this reduced system whose normal
form is of the same type as (\ref{v0}). The jump
of the active points after blow up is found by
solving the reduced system with a matching
condition \cite{else}. As we approach the
continuum limit, more and more points enter the
reduced system of equations and exponential
asymptotic methods become a viable alternative to
our methods.

In conclusion, we have studied depinning of
wavefronts in discrete RD equations. The normal
depinning transition can be viewed as a loss of
continuity of traveling front profiles as the
critical field is approached: below the critical
field, the fronts become pinned stationary
profiles with discontinuous jumps at discrete
values $u_n$. In the strongly discrete limit, the
critical field and these fronts can be
approximated by singular perturbation methods
which show excellent agreement with numerical
solutions. The leading order approximation to
the wavefront velocity is then correctly given
(scaling and prefactor) near the critical field.
Depinning transitions for discrete RD equations
apparently belong to two different universality
classes. In the normal class, the wavefront
velocity has a critical exponent 1/2. For certain
nonlinearities, the stationary fronts are
continuous functions of the discrete index at
zero field. Then the critical field is zero, the
depinning transition between stationary and
moving fronts is continuous, with a critical
exponent 1. This situation is the same as for
continuous RD equations and we have called it
anomalous pinning.

AC thanks S. Hastings and J.B. McLeod for fruifful
discussions.

\end{multicols}

\begin{references}{}
\bibitem{comparison}
If we have $l_n(0)\leq u_n(0)$ such that
$\dot{u}_n \geq u_{n+1}-2u_n+u_{n-1} + F -A
g(u_n)$, and $\dot{l}_n \leq l_{n+1}-2l_n+l_{n-1}
+ F -A g(l_n)$, then $l_n(t)\leq u_n(t)$ for all
later times. $l_n(t)$ and $u_n(t)$ are called sub
and supersolutions, respectively. See Ref.
\onlinecite{car99}.
\end{references}
\end{document}